\documentclass[a4paper,11pt]{article}
\pdfoutput=1 % if your are submitting a pdflatex (i.e. if you have
\usepackage{jheppub} % for details on the use of the package, please
\usepackage[T1]{fontenc} % if needed

\usepackage{natbib}
\usepackage{cancel}
\usepackage{enumitem}    
\usepackage{amsthm}         
\usepackage{amsmath} 	   
\usepackage{amssymb}       
\usepackage{amsfonts}
\usepackage{graphicx} 	    
\usepackage{verbatim}
\usepackage{color}
\usepackage{colortbl}
\usepackage{float}
\usepackage{multirow}
\definecolor{babyblue}{rgb}{0.54, 0.81, 0.94}
\definecolor{corn}{rgb}{0.98, 0.93, 0.36}

\title{\boldmath Cyclic completion of the anamorphic universe}

\author{Anna Ijjas}
\affiliation{Columbia Center for Theoretical Physics, New York, NY, 10027, USA}

\emailAdd{Anna.Ijjas@Columbia.edu}

\abstract{
Cyclic models of the universe have the advantage of avoiding initial conditions problems related to postulating any sort of beginning in time. To date, the best known viable examples of cyclic models have been ekpyrotic. In this paper, we show that the recently proposed anamorphic scenario can also be made cyclic.  The key to the cyclic completion is a classically stable, non-singular bounce. Remarkably, even though the bounce construction was originally developed to connect a period of contraction with a period of expansion both described by Einstein gravity, we show here that it can naturally be modified to connect an ordinary contracting phase described by Einstein gravity with a phase of anamorphic smoothing.
The paper will present the basic principles and steps in constructing cyclic anamorphic models.
}

\begin{document}
\maketitle
\flushbottom

%\noindent{\it Keywords\/}: anamorphic cosmology, null energy condition, non-singular Horndeski bounce
%\submitto{\cqg}

\section{Motivation}
\label{sec:secIntro}

It is the challenge of primordial cosmology to find a dynamical explanation for the initial conditions of the very early universe that led to the large-scale structure we observe today. Such an explanation is called for because the observed initial conditions appear to be highly tuned within standard hot big-bang cosmology: According to classical general relativity, there is no reason to expect that a slowly expanding patch of space would grow into a cosmological background that is flat and smooth on large scales and has nearly scale-invariant and gaussian density fluctuations. Even if we assumed initial homogeneity and flatness, decelerated expansion would quickly let inhomogeneities and spatial curvature grow and, hence, turn the initial patch of space into a `mess' not well suited for the evolution of the hierarchy of planets, stars, and galaxies we observe today.

Inflationary cosmology attempted to resolve the initial conditions problem by providing a mechanism insensitive to the physics of the big bang, {\it i.e.}, accelerated expansion was supposed to flatten and smooth the cosmological background and stretch quantum-generated density fluctuations over super-horizon scales to seed structure in the late universe \cite{Guth:1980zm,Albrecht:1982wi,Linde:1981mu,Bardeen:1983qw}. 
It has, though, been realized early on that the condition required for inflation to start -- a Hubble-sized, homogeneous patch dominated by the potential energy of a scalar field -- is rather delicate \cite{Penrose:1988mg,Gibbons:2006pa,Berezhiani:2015ola}. 
Furthermore, it is by now commonly accepted that inflation is generically eternal, implying that most of space-time is inflating and the volume of regions that are no longer inflating is measure zero \cite{Steinhardt:1982kg,Vilenkin:1983xq,Guth:2000ka}. Patches that complete inflation do so at different times after different random quantum fluctuations have affected the trajectory of the inflaton field that controls the rate of inflation. The different fluctuations lead to different cosmological outcomes. 
In addition, it is worth noting that the initial state at and immediately after the big bang is commonly assumed to be a highly energetic entangled quantum state. Although it is the minimal initial condition that any successful early-universe theory has to explain, we do not know any mechanism that would turn an initial quantum state into a classical space-time patch.
Inflation has not been designed to remove the classicality assumption either and, hence, cannot explain the quantum-to-classical transition, by construction. 

Cosmological scenarios proposing a contracting smoothing phase, such as the ekpyrotic universe \cite{Khoury:2001bz}, that is connected to the expanding phase of standard hot big-bang cosmology through a bounce successfully evade the `multiverse problem.' 
The multiverse problem arises when quantum fluctuations during smoothing lead to patches that explore an infinite variety of cosmological properties and there is no selection rule for deciding which is typical ({\it i.e.}, what numbers to expect in measuring these properties). 
The multiverse problem is generic in inflation, as pointed out by Linde and Guth (see, e.g., \cite{Guth:2007ng,Linde:2014nna}) although there exist specially designed potentials (see, {\it e.g.} \cite{Mukhanov:2014uwa}) that avoid it. We also note that, in models that match observations, inflation and the multiverse typically occur when the inflaton traverses one or more Planck units, which may conflict with UV completion; some have suggested this makes these ideas speculative.
By contrast, the anamorphic smoothing phase {\it generically} avoids the multiverse problem without adding any special features. 

The simple reason lies in the background behavior: Similar to inflation, rare adiabatic fluctuations can cause a patch to stay longer in the smoothing phase. However, now smoothing is achieved through contraction. Patches that contract longer reach the bounce later and start expanding later. Hence, eventually, patches delayed by rare fluctuations occupy exponentially less volume than the typical patches predicted by semi-classical physics and no multiverse problem arises.  
Also note that, unlike inflation, smoothing contraction begins when the universe is large and well-described by known semi-classical physics so that the classicality assumption at the beginning of the smoothing phase is  justified. Although there can arise many bubbles in the preceding dark-energy phase, the bubbles all undergo the same smoothing and thus have the same cosmological properties. 

Of course, an initial contracting phase does not resolve the initial conditions problem altogether. For a simple, `one-time bounce' scenario that begins with a contracting phase and transits to the current expanding phase with a bounce and ends with a phase dominated by dark energy, one still has to assume certain initial conditions to start smoothing contraction.
However, completing a contracting primordial scenario by introducing cycles of expansions followed by contraction  removes the initial conditions problem related to assuming a particular initial state a finite (physical) time ago.  
Furthermore, in contrast to inflation \cite{Borde:2001nh}, a cyclic scenario can be made geodesically complete and thereby avoid any  issues that arise in cosmological models associated with beginning the universe a finite (conformal) time ago  \cite{Bars:2013vba,Bars:2013qna}. 
Thus far, the classic ekpyrotic universe had been the only contracting scenario that has a cyclic completion \cite{Steinhardt:2002ih}.  It is nevertheless a commonly made mistake to use the notion `ekpyrotic' and `cyclic' as synonyms. For example, the `new ekpyrotic universe' introduced in \cite{Buchbinder:2007ad} is a one-time bounce scenario, and  here we will describe a cyclic scenario that is not ekpyrotic.

In this paper, we will present a particular realization of the recently proposed anamorphic scenario \cite{Ijjas:2015zma} and show that it can be made cyclic, featuring periods of contraction and expansion. 
Anamorphic cosmology is a novel approach to explain the smoothness
and flatness of the universe on large scales and the generation of a nearly scale-invariant spectrum of adiabatic density perturbations. The defining feature is a smoothing phase that acts like a contracting universe based on some Weyl frame-invariant criteria and an expanding
universe based on other frame-invariant criteria. An advantage of the contracting aspects is that it is possible to avoid the multiverse and measure problems that arise in inflationary models. Unlike ekpyrotic models, anamorphic models can be constructed with only a single field and can generate a nearly scale-invariant spectrum of tensor perturbations.
The key to the cyclic completion of anamorphic cosmology will be a stable, non-singular bounce. Originally, the bounce was developed to connect a phase of ordinary contraction to the current phase of expansion, both described by conventional Einstein gravity. Here, we will show that it is natural to modify the bounce to connect a phase of ordinary contraction with anamorphic smoothing. We will refer to this as the `$\Theta_{\rm Pl}$-bounce.'

The paper is organized as follows:
In Sec.~\ref{sec:secBasics}, we review the essentials of anamorphic cosmology including a novel Weyl-invariant formalism that enables us to track the cosmological evolution without reference to any particular frames. We outline the basic components of the cyclic scenario in Sec~\ref{sec:secSet-up}. The following Secs.~\ref{sec:secDE}-\ref{sec:secExit} describe the different cosmic stages of evolution: Starting from a dark-energy dominated phase, we show how it is possible for the universe to roll or tunnel to a short phase of  ordinary contraction that ends with a stable, non-singular $\Theta_{\rm Pl}$-bounce.
Remarkably, unlike in the cyclic ekpyrotic scenario, where the bounce connects the end of contraction with standard hot big-bang expansion, we show here how the same principles can be adapted to construct a non-singular $\Theta_{\rm Pl}$-bounce that generates the conditions required to begin a stage of anamorphic contraction. During the anamorphic phase, the cosmological background is smoothed and flattened and nearly scale-invariant, non-gaussian adiabatic modes are generated and stretched over super-horizon scales. After the smoothing phase ends, a non-singular $\Theta_m$-bounce connects to the expanding phase of standard hot big-bang evolution. We summarize the results and comment in Sec.~\ref{sec:secDiscussion}.

\section{Basics of anamorphic cosmology}
\label{sec:secBasics}

Anamorphic cosmology \cite{Ijjas:2015zma} has been proposed to  explain the smoothness and flatness of the universe on large scales and the generation of a nearly scale-invariant and gaussian spectrum of squeezed adiabatic density perturbations. 
Its defining feature is a smoothing phase in which the mass $m$ of massive particles and the Planck mass $M_{\rm Pl}$ have different time-dependence such that, relative to Compton wavelength of matter ($m^{-1}$), the smoothing phase acts like a contracting universe and, relative to the Planck length, it acts like an expanding universe.  For simplicity, we shall consider the case where matter-radiation consists of massive dust and the action for a single particle is $S_p = \int m\, ds$, where $ds$ is the line element and $m$ may vary with time.

If the particle mass and Planck mass were both constant, then the expansion or contraction of the physical background and the wavelengths of cosmological perturbations in  a homogeneous and isotropic Friedman-Robertson-Walker (FRW) universe (defined by the metric $ds^2 = - dt^2 + a^2(t)dx_i dx^i$, where $a(t)$ is the scale factor) are both unambiguously characterized by a a single quantity,  the scale factor $a(t)$, whose time variation is described by the Hubble parameter $H\equiv \dot{a}/a$.  However, during an anamorphic phase, when one or both masses are changing with time, it is useful to introduce two different measures: one, which measures  expansion or contraction  with respect to the particle Compton wavelength, is most useful for characterizing the expansion or contraction of the physical background; and the other, which measures expansion or contraction with respect to the Planck length, is most direct for characterizing the spectrum of cosmological perturbations.  

A dimensionless quantity that describes the physical expansion or contraction of the cosmological background
as measured relative to a ruler (or any object made of matter) is given by
\begin{equation}
\label{thetam}
\Theta_m = M_{\rm Pl}^{-1}\left( H + \frac{\dot{m}}{m} \right)
.
\end{equation}
The corresponding dimensionless quantity that measures the evolution
relative to the Planck mass, that determines the spectrum of scalar and tensor metric perturbations is given by
\begin{equation}
\Theta_{\rm Pl} = M_{\rm Pl}^{-1} \left( H + \frac{\dot{M}_{\rm Pl}}{M_{\rm Pl}} \right)
.
\end{equation}
Notably, the Hubble-like parameters $\Theta_{\rm Pl}$ and $\Theta_m$ are Weyl-frame independent. Using these parameters, different stages of cosmological scenarios can be unambiguously identified: 
In the cases of inflation, $\Theta_m = \Theta_{\rm Pl}>0$, and ekpyrosis,  $\Theta_m = \Theta_{\rm Pl}<0$, and in the present universe both signs are positive.  
The defining feature of anamorphic cosmology can be re-expressed as the requirement that $\Theta_{\rm Pl}$ and $\Theta_m$ have opposite signs; 
\begin{equation}
\Theta_m <0 \quad {\rm and} \quad \Theta_{\rm Pl}>0\,.
\end{equation} 
The condition in this equation will only apply during the anamorphic phase; in a complete cosmological model, the particle and Planck masses will become constant after the anamorphic smoothing phase (well before nucleosynthesis) and the condition $\Theta_m=\Theta_{\rm Pl}= (M_{\rm Pl}^0)^{-1} H>0$ will be reached and maintained through the present epoch, in accord with observations. Here, $M_{\rm Pl}^0 = 1/\sqrt{8\pi G_{\rm N}}$,  the current value of the reduced Planck mass and $G_{\rm N}$ is Newton's constant. 

To smooth and flatten the cosmological background on large enough scales during an anamorphic contracting phase, it is necessary that the anamorphic energy density dominates all other contributions to the total energy density for a sufficiently long time ($\sim 60$ $e$-folds of contraction).
The evolution of different forms of energy density  and curvature that contribute to the rate of contraction during the anamorphic phase is described by the first Friedmann equation, expressed in a frame-invariant form using $\Theta_m$:
\begin{equation}
\label{FriedmannEq1}
3\,  \Theta_m^2 \left(1 -  \frac{ d\ln \left( m/M_{\rm Pl} \right) }{ d\ln \alpha_m} \right)^2
= \frac{ \rho_{\rm A} }{ M_{\rm Pl}^4 }
+ \frac{\rho_m}{M_{\rm Pl}^4 }
- \left(\frac{m}{M_{\rm Pl}}\right)^2 \frac{\kappa}{\alpha_m^2} +  \left(\frac{m}{ M_{\rm Pl} }\right)^6 \frac{\sigma^2}{\alpha_m^6}.
\end{equation}
Here, the effective scale factor $\alpha_m$ is defined through $ \Theta_m \equiv M_{\rm Pl}^{-1}( \dot{\alpha}_m/\alpha_m) $.
The right hand side describes the different contributions to the total energy density:  $\rho_{\rm A}/M_{\rm Pl}^4 $, is due to the anamorphic energy density; $\rho_{\rm matter}/M_{\rm Pl}^4$, is due to the matter-radiation energy density with the matter consisting of particles with mass $m$; and the last two contributions are due to the spatial curvature, with $\kappa = (+1,0,-1)$,  and due to the anisotropy, parameterized by $\sigma^2$. Note that $\Theta_m < 0 < \Theta_{\rm Pl}$ implies that $M_{\rm Pl}^{-1}{\rm d}(m/M_{\rm Pl})/{\rm d}t \propto \Theta_m -\Theta_{\rm Pl}<0$; hence, in order for the anamorphic condition $\Theta_m < 0 < \Theta_{\rm Pl}$ to be satisfied, it is necessary that the invariant mass ratio $m/ M_{\rm Pl}$ decrease with time.
As a result, the factors of $m/M_{\rm Pl}$ in the Friedmann equation suppress the spatial curvature and, even more so, the anisotropy that would otherwise grow with shrinking $\alpha_m$. This is a key advantage of anamorphic contraction because suppressing the anisotropy in a contracting universe is essential to avoid chaotic mixmaster behavior while preserving smoothness and flatness \cite{Erickson:2003zm}. 

Defining the mass-variation index $q$ according to $1-q = {\rm d}\alpha_{\rm Pl}/{\rm d}\alpha_m$, it follows from ${\rm d}\alpha_{\rm Pl}/{\rm d}\alpha_m\propto \Theta_{\rm Pl}/\Theta_m <0$ that $q$ must be greater than 1. 
Neglecting the weak time-dependence of $q$, the smoothing condition that the anamorphic energy density dominates all other contributions on the right hand side of Eq.~\eqref{FriedmannEq1} reduces the Friedmann equation to the simple relation
\begin{equation}
\label{smoothing}
\Theta_m^2 (1-q)^2 \propto  \rho_{\rm A}/M_{\rm Pl}^4 \propto 1/ \alpha_m^{ 2 \epsilon_m },
\end{equation}
 where  we introduced the equation-of-state parameter 
\begin{equation}
\label{def-epsm}
\epsilon_m= -\frac{d\ln\Theta_m}{d\ln\alpha_m} \,.
\end{equation} 
In order for Eq.~\eqref{smoothing} to hold during contraction as $\alpha_m$ shrinks, $2\epsilon_m $ must exceed the corresponding exponents for the spatial curvature and anisotropy terms in the Friedmann equation~(\ref{FriedmannEq1}) when they are expressed as powers of $1/\alpha_m$.  
This condition yields a pair of constraints on $\epsilon_m$:
\begin{equation}
\label{smcon}
 \epsilon_m   \gtrsim 1 - q \quad \& \quad \epsilon_m   \gtrsim 3(1-q). 
\end{equation} 
With $q > 1$, both conditions are satisfied if the first inequality is satisfied, {\it i.e.}, the anamorphic smoothing condition is $ \epsilon_m   \gtrsim 1 - q $.   

For the generation of a nearly scale-invariant and gaussian spectrum of super-horizon adiabatic perturbations, the cosmological background must have the property that modes whose wavelengths are inside the horizon at the beginning of the smoothing phase can have wavelengths larger than the horizon size by the end of the smoothing phase. This is known as the `squeezing' condition.  The `horizon' is a dynamical length scale that separates smaller wavelengths for which the curvature modes are oscillatory from the large wavelengths for which the curvature modes become frozen.  In anamorphic models, the evolution of metric perturbations is fully determined  by the gravitational and scalar sector of the effective action; in particular, it does not depend on particle mass.  Accordingly, the dynamical length scale is set by $\Theta_{\rm Pl}^{-1}$ and the corresponding squeezing condition  
is that $\alpha_{\rm Pl} \Theta_{\rm Pl} $ be increasing,  
\begin{equation}
\label{sq1}
\frac{d |\alpha_{\rm Pl}\Theta_{\rm Pl}|}{d\,t} M_{\rm Pl}^{-1}  > 0,
\end{equation}
which reduces to the standard condition in inflationary and ekpyrotic models. The effective scale factor $\alpha_{\rm Pl}$ is given through $ \Theta_{\rm Pl} \equiv M_{\rm Pl}^{-1}( \dot{\alpha}_{\rm Pl}/\alpha_{\rm Pl}) $.
As shown in \cite{Ijjas:2015zma}, 
 the squeezing constraint in Eq.~\eqref{sq1} reduces to the same condition as the smoothing constraint in Eq.~\eqref{smcon} and, 
hence, squeezing imposes no additional constraint.
Due to the fact that $\Theta_{\rm Pl}>0$, the second-order action describing the generation and evolution of curvature perturbations during the anamorphic smoothing phase is similar to the case of inflation,
\begin{equation}
\label{S2-0}
S_{\zeta}^{(2)} = \int d^4 x\, \alpha_{\rm Pl}^3\,\epsilon_{\rm Pl}\left(\left(\frac{\dot{\zeta}}{M_{\rm Pl}}\right)^2 - \frac{c_S^2}{\alpha_{\rm Pl}^2}\left(\frac{\partial_i\zeta}{M_{\rm Pl}}\right)^2 \right),
\end{equation}
where $\zeta$ is the (frame-invariant) co-moving curvature perturbations given through the perturbed spatial metric $\delta g_{ij}=a(t) \exp(2\zeta({\bf x}, t))$; $c_S^2$ is the sound speed of co-moving curvature modes; and $\epsilon_{\rm Pl}$ is the equation of state parameter associated with $\Theta_{\rm Pl}$ via 
\begin{equation}
\label{def-epsPl}
\epsilon_{\rm Pl}= -\frac{d\ln\Theta_{\rm Pl}}{d\ln\alpha_{\rm Pl}} \,;
\end{equation} 
$\epsilon_{\rm Pl}\simeq$ constant during the anamorphic smoothing phase. 
Consequently, a nearly scale-invariant and gaussian spectrum of adiabatic curvature and tensor perturbations can be generated in anamorphic models with a single scalar field.
 
Obviously, anamorphic cosmology combines elements of earlier approaches, though, it does so in a novel way: Like in the ekpyrotic scenario, smoothing and flattening of the cosmological background is achieved through contraction and the bang is replaced by a bounce that connects the anamorphic contracting phase with the expanding hot big-bang phase.  
On the other hand, during an anamorphic phase, it is possible to generate (nearly) scale-invariant, squeezed adiabatic perturbations with only a single scalar field. During an ekpyrotic phase, curvature perturbations decay \cite{Creminelli:2004jg} so that one typically invokes a two-field entropic mechanism to first generate squeezed entropy modes and later convert them to adiabatic modes after the ekpyrotic phase ends, before or during the bounce \cite{Lehners:2007ac,Buchbinder:2007ad}. 
Like in inflation, as measured relative to the Planck scale,  the anamorphic phase resembles expansion and thus curvature perturbations grow and become squeezed. But eternal smoothing and the multiverse problem are avoided due to contraction of the physical background.  

Notably, to connect the anamorphic smoothing phase with the hot expanding phase of standard big-bang evolution, there is a simple, `anamorphic way' to have a one-time bounce:  For example, as has been shown in \cite{Ijjas:2015zma}, scalar-tensor realizations of anamorphic cosmology allow for the Hubble-like parameter $\Theta_m$ to increase with time and eventually hit zero ($\Theta_m$-bounce) while $\Theta_{\rm Pl}>0$ throughout cosmic evolution, all without encountering instabilities.
However, as discussed above, a one-time scenario that starts with an anamorphic smoothing phase and connects to standard hot big-bang evolution through a simple $\Theta_m$-bounce does not remove the initial conditions problem altogether.
In the remainder of this paper, we will present a particular example of the anamorphic scenario and show that it can be made cyclic, resolving issues related to assuming any sort of beginning in time. 

\section{Essential ingredients}
\label{sec:secSet-up}

We will analyze a particular microphysical realization of the anamorphic theory involving a single scalar field $\phi$ given by the  action
\begin{align}
\label{JF-action1}
S &= \int d^4x \sqrt{-g} \Bigg( \frac{1}{2}M_{\rm Pl}^2(\phi) R -  \frac{1}{2} K(\phi) (\partial\phi)^2 - \left(\frac{m_0}{m(\phi)}\right)^4V(\phi) \Bigg) 
+ S_{\rm matter} 
\\
&+ [{\rm terms\; only\; important\; near\; the\;} \Theta_{\rm Pl}{\rm-bounce;\; see\;Sec.~\ref{sec:secNEC1}}]\nonumber
,
\end{align}
where $g$ is the metric determinant; $R$ is the Ricci scalar; and the quadratic coupling is defined through
\begin{equation}
\label{def-K}
K(\phi) = \left(\frac{m_0}{m(\phi)}\right)^2 \left(k(\phi)+6\frac{m_0^2m,^2_{\phi}}{m^4(\phi)}\right);
\end{equation}
note that $K(\phi)=k(\phi)$ in the representation where $\dot{m}=0$ and thus $k(\phi)$ cannot simply be absorbed into the definition of $\phi$ because it goes through zero during the cycle, as we will show below. 
The action is an example of generalized Brans-Dicke theories; it  describes a single scalar field $\phi$ non-minimally coupled to gravity, where 
$M_{\rm Pl}^2(\phi)$ is the effective (reduced) Planck mass and $m$ is the test-particle mass, as defined above. Before and after the anamorphic smoothing phase both $m$ and $M_{\rm Pl}$ are constant, with  $M_{\rm Pl}^2=(M_{\rm Pl}^0)^2$,  the current value of the reduced Planck mass, and $m=m_0$, the current value of the unit test-particle mass. Hence, all standard tests of Einstein gravity are satisfied since the anamorphic phase is over well before nucleosynthesis.  Throughout,  we work in natural units $M_{\rm Pl}^0\equiv1$ unless otherwise noted.
 The field's potential energy is characterized by $V(\phi)$. 
As per convention, we call the representation of the action where the matter Lagrangian $\mathcal{L}_m$ is independent of the scalar field $\phi$, or equivalently, $m\equiv m_0$ the `Jordan-frame representation.' The `Einstein-frame' where the gravitational sector is independent of the scalar and matter is non-trivially coupled to $\phi$, or equivalently, $M_{\rm Pl}\equiv M_{\rm Pl}^0$ is related to the Jordan frame through a conformal transformation $g_{\mu\nu}\to \left(M_{\rm Pl}/M_{\rm Pl}^0\right)^2 g_{\mu\nu}$. Whenever it is possible, we will employ a frame-independent analysis. 

In a flat FRW universe, the corresponding  background equations are given by
\begin{align}
\label{JF1}
3\Theta_m^2  & =
\frac{1}{2} k(\phi) \frac{\dot{\phi}^2}{M_{\rm Pl}^2} + \left(m_0^{-1}\frac{m}{M_{\rm Pl}}\right)^2\left( 6\Theta_m \frac{d}{M_{\rm Pl}dt}\ln\left(m_0^{-1}\frac{m}{M_{\rm Pl}}\right) + V(\phi)\right)\,,\\
\label{JF2}
-\frac{\dot{\Theta}_m}{M_{\rm Pl}}   &= \left(m_0^{-1}\frac{m}{M_{\rm Pl}}\right)^2 \left( \frac{1}{2} k(\phi)+ \left(m_0^{-1} \frac{d}{d\phi}\frac{m}{M_{\rm Pl}} \right)^2 \right)\frac{\dot{\phi}^2}{M_{\rm Pl}^2}   -  2\Theta_m \frac{d}{M_{\rm Pl}dt}\ln\left(m_0^{-1}\frac{m}{M_{\rm Pl}}\right) \qquad 
\nonumber \\
&- \left(\frac{\ddot{\phi}}{M_{\rm Pl}^2} + \frac{d}{M_{\rm Pl}^2dt}\ln\left(\frac{m_0}{m}\right)\right)\frac{d}{d\phi}\ln\left(m_0^{-1}\frac{m}{M_{\rm Pl}}\right)  - \frac{\dot{\phi}^2}{M_{\rm Pl}^2}\frac{d^2}{d\phi^2}\ln\left(m_0^{-1}\frac{m}{M_{\rm Pl}}\right)
\,,
\end{align}
and, as usual, the scalar-field equation can be easily obtained by combining the time derivative of the first Friedmann equation with the second Friedmann equation.
Here, the frame-independent Hubble-like parameter $\Theta_m$ characterizes the behavior of the physical background ( $3\Theta_m^2=\rho/(m_0^2M_{\rm Pl}^2)$) as measured by rulers made out of matter and the test-particle mass $m$, as defined above in Sec.~\ref{sec:secBasics}.

%%%%%FIGURE~1
\begin{figure}[tb]
\begin{center}
\includegraphics[width=12.5cm]{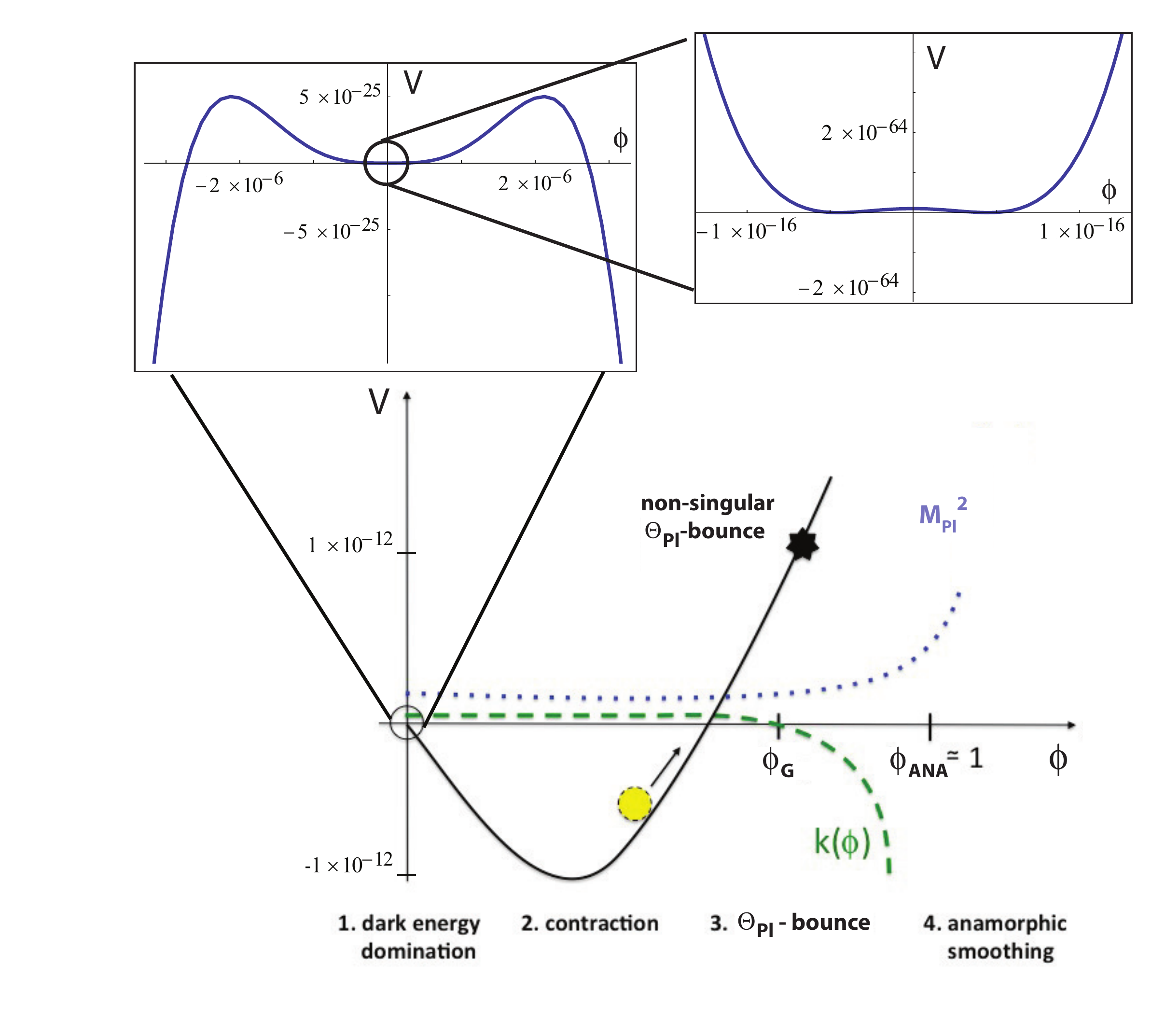}
\caption{A schematic illustration of the scalar potential $V$ (continuous black line)  as a function of the anamorphic field $\phi$ in the cyclic anamorphic scenario with the middle inset showing the energy barrier and the final inset showing the current metastable vacuum. Both the anamorphic field $\phi$ and its potential $V$ are given in Planck units. We superposed curves to represent the gravitational coupling, $M^2_{\rm Pl}$ (dotted-dashed blue line) and the kinetic coupling, $k$ (dashed green line) that both are dimensionless functions of $\phi$. For small field values ($\phi\ll \phi_{\rm ANA}$) and, in particular, during the current cosmological epoch when the anamorphic field is trapped in the metastable vacuum, $M^2_{\rm Pl}= 1$ and $k=1$, in agreement with observations; the plots of $M^2_{\rm Pl}$ and $k$ have been shifted apart for the purpose of illustration.
}
\label{FigCompleteScenario}
\end{center}
\end{figure}

The various terms in the action and background equations play different roles at different cosmological stages each imposing constraints on the associated couplings. The key phases of a complete cycle are illustrated in Fig.~\ref{FigCompleteScenario}:
During the current dark energy-dominated phase ($\phi=\phi_0\simeq 10^{-16} M_{\rm Pl}^0\ll \phi_{\rm ANA}$), the anamorphic field is trapped in a vacuum state, the effective Planck mass as well as the test-particle mass are constant ($M_{\rm Pl}(\phi_0)=M_{\rm Pl}^0$, $m(\phi_0)=m_0$) so that $H=\Theta_{\rm Pl}=\Theta_m>0$. For the anamorphic smoothing phase to begin ($\phi \sim \phi_{\rm ANA} $), on the other hand,
the field must be high up its potential energy curve and the Hubble-like parameters must have opposite signs with $\Theta_m<0<\Theta_{\rm Pl}$.

For a cyclic completion of the anamorphic scenario, we have to connect the current phase with the anamorphic phase, setting the initial conditions for smoothing contraction; and, after the smoothing phase ends, the universe has to settle in a metastable vacuum state. In the forthcoming
sections, we shall describe how each stage of cosmic evolution transits to the next phase, interweaving a combination of concepts:
\begin{enumerate}
\item[] {\it Stage 1: Current vacuum.} The anamorphic field is settled in a low energy density, metastable vacuum state. The dark energy-dominated phase ends by the field tunneling or rolling to a state with negative potential energy density. (Sec.~\ref{sec:secDE})
\item[]{\it Stage 2: Ordinary contraction.} Since the true vacuum energy is negative, a period of contraction begins that drives the field uphill in the potential to approximately the value of $V(\phi)$ needed to generate energy-density perturbations of the right amplitude. During this phase, $M_{\rm Pl}=M_{\rm Pl}^0$ and $m=m_0$ such that $H =\Theta_{\rm Pl}=\Theta_m$ and the evolution is described by conventional Einstein gravity. (Sec.~\ref{sec:secDE}).
\item[]{\it Stage 3: $\Theta_{\rm Pl}$-bounce.} 
When the field is high up the potential, the universe enters a brief period that flips the sign of $\Theta_{\rm Pl}$ to be positive and opposite that of $\Theta_m$ ($\Theta_{\rm Pl}$-bounce). The $\Theta_{\rm Pl}$-bounce is achieved by modifying our recently proposed, stable non-singular bounce construction to connect a phase of ordinary contraction with a phase of anamorphic smoothing. This stage is the new addition to the anamorphic scenario and the key to the cyclic completion of the anamorphic scenario. (Sec.~\ref{sec:thetaPl-bounce})
 \item[] {\it Stage 4: Anamorphic smoothing.} 
 During the anamorphic smoothing phase, the cosmological background is smoothed and flattened and nearly scale-invariant, adiabatic and gaussian modes are generated and stretched over super-horizon scales. (Sec.~\ref{sec:secExit})
  \item[] {\it Stage 5: $\Theta_m$-bounce to start hot big bang evolution} 
The anamorphic smoothing phase ends in a finite time when $\phi$ reaches $-\infty$. During the following phase of slow expansion the anamorphic field rolls back to and settles  in the metastable vacuum state where the cycle restarts (Sec.~\ref{sec:secExit}). 
\end{enumerate}
%%%%%%%%%%%%%%

\section{From dark-energy domination to (ordinary) contraction}
\label{sec:secDE}

As a concrete example, let's assume that the potential for the anamorphic field has the form shown in Fig.~\ref{FigCompleteScenario}. Let's further assume that today's vacuum  is metastable and $\phi_0$ has the observed value $\phi_0\simeq 10^{-16} M_{\rm Pl}^0$, comparable to the Higgs' expectation value.     That means, in the current cosmological phase, the anamorphic field is settled in the local, metastable vacuum state of a Higgs-like potential,
\begin{equation}
\label{V1}
V\sim - \mu^2\phi^2 + \lambda(\phi)\phi^4
\,,
\end{equation}
in which $\lambda$ runs from positive to negative values between $\phi_0\leq\phi\leq\phi_{\rm min}$, where $\phi_{\rm min}$ is the value at the absolute minimum of the potential $-V_{\rm min}$. The metastable vacuum at $\phi=\phi_0$ is separated by a barrier of $\sim (10^{10-12}\, \mathrm{GeV})^4$ from a true, negative energy-density vacuum at $\phi=\phi_{\rm min}$. (Note that the potential beyond the barrier has an absolute minimum described below and is not meant to be that of the actual Higgs; we only use this example for illustration.)
The potential energy density $V(\phi_0)$ is the dominant energy component and it has the observed value of today's vacuum or dark-energy density $\sim (10^{-12}\, \mathrm{GeV})^4$; the gravitational coupling is $M_{\rm Pl}(\phi_0) \equiv 1$ (in agreement with current tests of general relativity) and the kinetic coupling is $k \equiv 1 $ (in accordance with the standard Higgs model).
Due to the positive vacuum energy density of the metastable vacuum, the cosmological background expands at an accelerated rate. 

Eventually, the metastable vacuum decays and the field ($10^{-6} M_{\rm Pl}^0 \lesssim \phi \lesssim \phi_G$) rolls or tunnels to the negative part of the potential,
\begin{equation}
\label{V2}
V\sim -\bar{\lambda} \phi^4 \,,
\end{equation}
where $-\bar{\lambda}<0$.
 As the field rolls downhill to more negative potential energy density, its kinetic energy grows while $H$ keeps decreasing ($\dot{H}=-\dot{\phi}^2/2$) so that $\Theta_m=\Theta_{\mathrm{Pl}}=H$; both undergo a reversal from positive to negative and the universe starts contracting.  During the contracting phase that follows, the blue shifting kinetic energy density rapidly dominates the total energy density such that  
 \begin{equation}
 \epsilon =  3\,\frac{ \frac{1}{2}\dot{\phi}^2 }{\frac{1}{2}\dot{\phi}^2+V}\gtrsim 3.
 \end{equation}
Here, $\epsilon$ is the equation of state parameter defined as $(3/2)(1+w)$ where $w=p/\rho$ with $\rho$ being the energy density and $p$ the pressure of the anamorphic field; the macrophysical quantities $\rho$ and $p$ relate to the canonical microphysical scalar as per convention, 
 \begin{align}
\rho &= \frac12 \dot{\phi}^2+V(\phi)\,,\\
p &= \frac12\dot{\phi}^2-V(\phi)
\,. 
\end{align}
Note that $\epsilon = \epsilon_m=\epsilon_{\rm Pl}$ if  $H=\Theta_m=\Theta_{\rm Pl}$, with $\epsilon_m, \epsilon_{\rm Pl}$ as defined in Eqs.~(\ref{def-epsm}-\ref{def-epsPl}).

 Rolling downhill, the anamorphic field keeps picking up kinetic energy and, due to its large kinetic energy density ($\epsilon \gg 3$) at the bottom of the potential, $V_{\rm min}<0$, the field does not settle in the true, negative vacuum ($\phi=\phi_{\rm min}$) but overshoots and continues to increase and roll uphill with a scalar potential that, for the purposes of illustration, we will take to be well-approximated by the (near) exponential
 \begin{equation}
\label{V3}
V\sim  -V_{\rm min} + V_0\exp\left( \sqrt{2\bar{\epsilon}} \left(\phi-\phi_{\rm NEC}\right)\right) \quad {\rm with}\quad {\bar{\epsilon}} \sim 3\,.
\end{equation}
While the field is rolling uphill, the background keeps contracting and admits the well-known scaling solution for exponential potentials, 
 \begin{equation}
a\sim(-t)^{1/ \bar{\epsilon}},\quad \phi \sim -\frac{2}{ \bar{\epsilon} }\ln\Big( V_0\bar{\epsilon}\,  (-t) \Big), \quad \phi_0=\phi(t_0)\,,
\end{equation}
where $t_0$ is given implicitly through $V(\phi_0)=0$ and $t$ is going from large negative to small negative numbers.
As it rolls uphill, the field's kinetic energy grows at a slower rate  but continues to dominate ($\epsilon \sim 3$)  for a substantial period after $V(\phi)$ changes from negative to positive.  Obviously, the parameters $V_{\rm min}$ and $V_0$ can be chosen such that,  for $\epsilon=\bar{\epsilon} \sim 3$,  
\begin{equation}
\delta\rho/\rho\sim V^{3/2}/V,_{\phi}\gtrsim 10^{-5}
\end{equation}
when $\phi = \phi_G$; {\it i.e.}, at $\phi = \phi_G$ the field is sufficiently high up the potential to begin generating the right spectrum of co-moving curvature perturbations.

\section{The $\Theta_{\text{Pl}}$-bounce stage}
\label{sec:secNEC1}

For the anamorphic smoothing phase to start, it remains to turn on the effective Planck mass $M_{\rm Pl}$ or, equivalently, the particle mass $m$ such that $\Theta_m=M_{\rm Pl}^{-1}\left(H+\dot{m}/m\right) < \Theta_{\rm Pl}=M_{\rm Pl}^{-1}\left( H+\dot{M}_{\rm Pl}/M_{\rm Pl}\right)$ and to flip the sign of $\Theta_{\rm Pl}$ from negative to positive ($\Theta_{\rm Pl}$-bounce). 

In order to flip the sign of the Hubble-like parameter $\Theta_{\rm Pl}$ from negative to positive ($\Theta_{\rm Pl}$-bounce), there has to be a brief phase where $\dot{\Theta}_{\rm Pl}>0$. However, to do this with an ordinary scalar, such as the anamorphic field would introduce pathologies. The pathology can be seen by examining the action describing the metric perturbations $\zeta$ given in  Eq.~\eqref{S2-0}.  From Eq.~\eqref{def-epsPl}, it follows that the kinetic coefficient of the modes $\epsilon_{\rm Pl}$ is directly proportional to the time derivative of ${\Theta}_{\rm Pl}$, 
\begin{equation}
\epsilon_{\rm Pl}=-M_{\rm Pl}^{-1}\frac{\dot{\Theta}_{\rm Pl}}{{\Theta}_{\rm Pl}^2}
\,.
\end{equation}
If $\dot{\Theta}_{\rm Pl}>0$ as required for the $\Theta_{\rm Pl}$-bounce, the anamorphic theory would suffer from a ghost instability. Notably, simply adding higher-order kinetic terms $\sim (\partial\phi)^{2n}$ (where $n>1$) removes the quantum ghost but introduces classical `gradient instabilities' \cite{Rubakov:2014jja}. More exactly, higher-order kinetic terms that remove the ghost in the action but lead to wrong-sign gradients spoil the character of the evolution equation altogether. If $\dot{\Theta}_{\rm Pl}>0$, these terms turn a strongly hyperbolic system into an elliptic one, such that the initial value problem becomes ill-posed,  {\it i.e.}, a generic linear perturbation cannot be evolved in time so dynamics no longer makes sense \cite{Papallo:2017qvl}. Most importantly, the wrong-sign gradient  term is a more serious issue than a dynamical instability that takes time to develop. The effect is instantaneous.  It is a misconception, for example, that an imaginary sound speed  can be tolerated for short periods of time because the `gradient instability' takes time to grow.\footnote{
There have been attempts to patch the ultra-high frequency modes by introducing higher-order operators $\sim (\Box \phi)^2$ \cite{Creminelli:2006xe,deRham:2017aoj}. Obviously, such terms lead to third derivatives in the equations of motion. The resulting additional propagating degree of freedom is commonly referred to as an Ostrogradsky ghost. The conventional argument is that a heavy ghost only gets excited in the UV and, hence, it would not influence the background evolution. But the smallness of these terms (or equivalently, a large mass for the ghost) is entirely irrelevant from a dynamical point of view. Once the terms are present in the action, no matter how small they are, the character of the evolution equations immediately changes and the theory becomes ill-posed so one cannot trust the presumed background solution at low energies.}

Consequently, for a stable sign-flip of $\Theta_{\rm Pl}$, we have to amend the anamorphic action in Eq.~\eqref{JF-action1} in a more sophisticated way that leads to a modification of the linearized action, allowing for a short period during which $\Theta_{\rm Pl}<0$ increases, changes sign and reaches approximately the  same magnitude (but positive) as before without introducing pathologies. (Note that the modified action will only be the effective description of cosmic evolution for a very short period during which the $\Theta_{\rm Pl}$-bounce takes place. Outside the $\Theta_{\rm Pl}$-bounce stage, the simple anamorphic action governs the evolution.)

To achieve such a stable $\Theta_{\rm Pl}$-bounce, setting the stage for anamorphic smoothing to start, we shall adapt the fully stable, classical non-singular bounce construction introduced in Refs.~\cite{Ijjas:2016tpn,Ijjas:2016vtq,Ijjas:2017pei}. In that construction, a phase of ordinary contraction was connected to a phase of ordinary expansion, both described by Einstein gravity.
Here there will be one important modification: the $\Theta_{\rm Pl}$-bounce will connect a phase of ordinary contraction described by Einstein gravity to a phase of anamorphic smoothing that includes modifications to Einstein gravity. To this end, in the remainder of the section, we first briefly review the results of Refs.~\cite{Ijjas:2016tpn,Ijjas:2016vtq,Ijjas:2017pei}, pointing out some fine points not noted previously, and then present a procedure for constructing a stable $\Theta_{\rm Pl}$-bounce. 

\subsection{Fully stable, non-singular Horndeski bounce: a recap}
\label{sec:recap}

In Ref.~\cite{Ijjas:2016tpn}, we have shown that it is possible to transit from a period of contraction ($H<0$) to a period of expansion ($H>0$) at low energies and finite values of the scale factor $a$ without encountering pathologies or singular behavior; {\it i.e.}, the universe can bounce. In addition, in Ref.~\cite{Ijjas:2016vtq} we presented fully stable cosmological solutions that involve a classical non-singular bounce; notably, while gravity has to be modified during the bounce stage, we gave examples where, after the bounce, cosmic evolution is described by conventional Einstein gravity.  Finally, in Ref.~\cite{Ijjas:2017pei} we demonstrated that a smooth transition from Einstein to modified gravity before the non-singular bounce stage is possible such that both before and after the bounce, cosmic evolution is described by Einstein gravity.

It is the characteristic feature of a classical, non-singular bounce that the Hubble parameter is increasing ($\dot{H}>0$) during the bounce stage. For an FRW cosmological background, this requires violation of the null energy condition (NEC).
The historical challenge to achieve a stable bounce stage was that, in practice, NEC-violation during the bounce stage seemed to either lead to ghost or gradient instabilities. Recently, Libanov et al. \cite{Libanov:2016kfc} and Kobayashi \cite{Kobayashi:2016xpl} each presented no-go theorems according to which, in the absence of ghost, NEC violation in theories involving a single scalar leads either to wrong-sign gradient terms or singularities.

For our stable bounce construction, we exploited a loophole of these no-go arguments: We used the same `ingredient' as considered in the no-goes --  an imperfect fluid component -- but have shown that its `braiding' with gravity in a non-trivial way indeed makes  avoiding pathologies possible during the non-singular bounce stage. 
The underlying microphysical theory is described by the conformally-coupled ${\cal L}_4$-Horndeski action \cite{Horndeski:1974wa},
\begin{equation}
\label{action0}
S = \int d^4x \sqrt{-g}\,\left(\frac{1}{2}R + \sum^4_{i=2}{\cal L}_i \right)+S_{\rm matter}\,,
\end{equation}
where the Lagrangian density of the scalar is given as the sum of the $i$th Horndeski Lagrangians
\begin{eqnarray}
\label{L0}
{\cal L}_2 &=& k(\phi)X + q(\phi)X^2 - V(\phi), \\
\label{L3}
{\cal L}_3 &=& -  b(\phi)X\Box\phi ,\quad \\
\label{L4}
{\cal L}_4 &=& \frac{1}{2} f(\phi)R 
\,.\nonumber
\end{eqnarray}
Here, $X= -(1/2)\partial_{\mu}\phi\partial^{\mu}\phi$ is the canonical kinetic term; $k(\phi)$ is the dimensionless quadratic coupling; $q(\phi)$ is the dimensionless quartic coupling;  $b(\phi)$ is the dimensionless coupling of the scalar field $\phi$ to the ${\cal L}_3$-Horndeski term; and $V(\phi)$ is the scalar potential. The positive definite, conformal coupling to the Ricci scalar $(1/2)f(\phi)$ is distinctive of the ${\cal L}_4$-Horndeski interaction. Notably, to date the conformally-coupled ${\cal L}_4$-Horndeski theory is the most general Lorentz-invariant theories that is, at the same time, proven to be (linearly) well-posed \cite{Papallo:2017qvl}. On the other hand, more general modifications of Einstein gravity appear to be ill-posed and os it is unclear how to extract actual predictions from these theories.  This feature makes the conformally-coupled ${\cal L}_4$-Horndeski theory particularly well-motivated for cosmological applications.

Varying the action in Eq.~\eqref{action0} yields the corresponding FRW background equations 
 \begin{align}
\label{EF1-FRW-G}
3 H^2  &=  \frac{1}{2}k(\phi)\dot{\phi}^2 + \frac{1}{4}\left(3 q(\phi) -2b,_{\phi}\right)\dot{\phi}^4 + V(\phi) + 3\left( b(\phi)\dot{\phi}^3 -f,_{\phi}\dot{\phi} \right) H  \\
& - 3f(\phi)H^2 
\,, \nonumber \\
\label{EF2-FRW-G}
-2\dot{H} & =  \left(k(\phi)+ f,_{\phi\phi}\right)\dot{\phi}^2 +  \left(q(\phi)-b,_{\phi} \right) \dot{\phi}^4 
+ \left( 3b(\phi) \dot{\phi}^3 - f,_{\phi}\dot{\phi} \right) H
\\
& - \left( b(\phi) \dot{\phi}^2 - f,_{\phi}  \right)\ddot{\phi}   + 2\dot{H}f(\phi) 
.\nonumber
\end{align}
Note that, due to the non-trivial `braiding' between gravity and the imperfect Horndeski scalar, the theory is not conformally (nor disformally) invariant to Einstein gravity. Consequently, the co-variant Horndeski action in Eq.~\eqref{action0} cannot be re-formulated in a way that the scalar and gravity sectors are separated and, hence, there is no `Einstein-frame representation.' Indeed, generalized Brans-Dicke theories, such as the anamorphic theory, are special examples of Horndeski theories (with $q, b \equiv 0$) where, due to the perfect-fluid form,  a separation of gravitational and field sectors is possible so that the definition of the Hubble-like invariants $\Theta_m, \Theta_{\rm Pl}$ is unambiguous. For the Horndeski scalar, on the other hand, $\Theta_m$ is only well-defined up to a positive definite normalization factor  ($\Theta_m \propto H$) and it is not possible to identify a corresponding $\Theta_{\rm Pl}$.   There is, though, no ambiguity whatsoever in identifying the behavior of the physical background. Since matter and radiation are decoupled from the scalar, a negative Hubble parameter ($\Theta_m \propto H<0$) means contraction and a positive Hubble parameter ($\Theta_m \propto H>0$) means expansion. The cosmological bounce occurs at $\Theta_m=H=0$.

%Perturbations
Due to the fact that the leading order instabilities source from linear perturbations, the stability behavior during the bounce stage can be checked using the linear theory. 
In Ref.~\cite{Ijjas:2016tpn}, we have shown that the unitary gauge ($\delta \phi =0$) is well-suited to analyze stability during the bounce stage ($\dot{H}>0$).
The second-order Horndeski action for the metric perturbation $\zeta$ is given by
\begin{equation}
\label{zeta-action}
S^{(2)}_{\zeta}=\int d^4x \,a^3 A(t) \left( \dot{\zeta}^2- c_{\zeta}^2(t)\frac{\left(\partial_i\zeta\right)^2}{a^2}\right)\,,
\end{equation}
where the ratio $c_{\zeta}^2=B/A$ defines the propagation speed of $\zeta$-perturbations and the coefficients of the kinetic and gradient terms are defined as
\begin{align}
\label{Aeq}
A(t) &= m_{\alpha}(t) \left(\frac{A_h(t)}{\gamma(t)}\right)^2 + 3A_h(t)
\,,\\
\label{Beq}
B(t) &= a^{-1}(t)\frac{d}{dt}\left(a(t)\frac{A_h^2(t)}{\gamma(t)}\right) - A_h(t)
\,;
\end{align}
with
\begin{align}
\label{ah0}
A_{h} &= 1+f(\phi) 
\,,\\
\label{malpha}
m_{\alpha}&=  \frac{1}{2}k(\phi) \dot{\phi}^2+\frac{1}{2}\left(3q(\phi) -2b,_{\phi}
\right)\dot{\phi}^4  -3f,_{\phi}H\dot{\phi}   + 6 b(\phi)   H\dot{\phi}^3
-  3\left( 1+f (\phi)  \right)H^2 
\,,\\
\label{gamma}
\gamma &= \left( 1+f(\phi) \right)H + \frac{1}{2}f,_{\phi}\dot{\phi} - \frac{1}{2}b(\phi)\dot{\phi}^3 
. 
\end{align}
Accordingly, for the $\zeta$-perturbation to pass through smoothly, the sound speed $c_{\zeta}^2$ has to be strictly positive throughout, with $A, B>0$.

Similarly, the second-order action for tensor modes takes the simple form
\begin{equation}
S^{(2)}_{h_{ij}}=\int d^4x\, a^3 A_h(t) \left( \dot{h}^2_{ij}- c_T^2(t)\frac{\left(\partial_l h_{ij}\right)^2}{a^2}\right)
\,,
\end{equation}
where $c_T^2\equiv 1$ is the squared sound speed of tensor modes. Obviously, due to the purely conformal coupling, tensor modes propagate stably as in conventional Einstein gravity.
%%%

The key to our bounce construction as presented in Ref.~\cite{Ijjas:2016tpn}  has been to show that a stable Horndeski bounce can be achieved if  the quantity $\gamma$ in Eq.~\eqref{gamma} that measures the strength of `braiding' between the scalar and the metric  keeps its sign throughout the NEC-violating bounce stage when $\dot{H}\geq0$, since avoiding a zero-crossing of $\gamma$ is equivalent to having $B(t)$ in Eq.~\eqref{Beq} remain positive and finite (what was believed to be impossible previously). 
For our examples, we employed an `inverse method,' specifying the functions $H, \gamma, V$ and $A_h$ as input to ensure stability and non-singular behavior throughout the bounce stage. Finally, fixing $\dot{\phi}$ as well,  we were able to recover the corresponding couplings $f(\phi), k(\phi), q(\phi)$ and $b(\phi)$ in terms of these functions:
\begin{align}
\label{f_1(t)}
f  &=  A_h -1 \,,
\\
\label{EF1-FRW-k}
k  &= -\frac{2}{\dot{\phi}^2} \bigg( \dot{\gamma} + 3H\gamma + 2\frac{d}{dt}\left( A_{h} H \right)
+ 3A_{h} H^2 +\ddot{A}_h -2V \bigg)
\,,
\\
\label{EF2-FRW-q}
q &= \frac{4}{3\dot{\phi}^4}\bigg(\dot{\gamma}+ 9H\gamma + 2\frac{d}{dt}\left( A_{h} H \right)  + \ddot{A}_h -3V \bigg) + \frac{2}{3}b,_{\phi}
\,,\\
\label{b(t)}
b &= \frac{2}{\dot{\phi}^3} \left(  A_h H - \gamma + \frac{1}{2}\dot{A}_h \right)
\,,
\end{align}
and show they are well-behaved for the given input functions.
Our inverse method makes it possible to search rapidly and systematically through different bouncing Horndeski models with a given background profile and identify the stable ones.
Indeed, a simple ${\cal L}_3$-Horndeski action (where $f(\phi) \equiv 0$) does the work. 

For fully stable cosmological solutions involving a non-singular bounce from a phase of contraction to a phase of expansion described by Einstein gravity (that is, including all times to $\pm \infty$ before and after the bounce stage), we have shown in Ref.~\cite{Ijjas:2016vtq} that it is necessary for $\gamma$ to switch sign before or after NEC violation. The reason is that smoothly connecting the Horndeski Lagrangian to ordinary Einstein gravity with a canonical scalar means that, before entering the bounce stage, $\gamma = H<0$; and, after exiting, $\gamma = H>0$. Most recently, in Ref.~\cite{Ijjas:2017pei}, we performed a linear stability analysis around `$\gamma$-crossing ($\gamma = 0$) and proved that all modes with wavelength larger than the Planck length pass through smoothly for generic backgrounds $\gamma\sim t-t_\gamma, H \sim 1/(-t)$ and $A_h \sim 1+ A_{\gamma}(t-t_\gamma)^2$. 
%Without the ${\cal L}_4$ terms, $B(t)$ would either blow up or become negative within a finite time before or after the NEC-violating bounce stage when $\gamma$ flips sign from positive to negative. It is easy to see, though, from Eq.~\eqref{Beq} that, with the ${\cal L}_4$-Horndeski,  it is possible to have $\gamma$ cross through zero, with $\gamma\sim t-t_\gamma$ in the proximity of the zero-crossing point $\gamma(t_{\gamma})=0$, without encountering a pathology provided  $A, B, A_h, B_h \sim (t-t_\gamma)^2$ and $c_S^2=B/A>0, (c_S^T)^2=B_h/A_h>0$.  

As a technical point, note that, as $\gamma$ approaches $0$, $\zeta\to\infty$; but, as we have shown in~\cite{Ijjas:2017pei}, this is a mere formal blow-up ({\it i.e.}, a `coordinate singularity') of a variable that is ill-suited to describe the physics near $\gamma=0$. This can be easily seen by choosing a different gauge, {\it e.g.}, Newtonian or harmonic. In fact, due to the braiding between the Horndeski scalar and the metric, $\zeta$ ceases to be the co-moving curvature perturbation. In particular, the criterion that $c_{\zeta}^2$, the sound speed of $\zeta$, be positive in ${\cal L}_3$ or more general Horndeski theories is sufficient but not a necessary condition for stability.  Indeed, in certain cases (such as around $\gamma$-crossing), demanding  $c_{\zeta}^2>0$ overconstrains the  theory.

\subsection{Stable $\Theta_{\rm Pl}$-bounce with no $\Theta_m$-bounce}
\label{sec:thetaPl-bounce}

The cosmological bounce presented in the previous section is a $\Theta_m$-bounce in the sense that it smoothly connects a contracting with an expanding physical background. Now we will show that we can modify this approach to construct a stable, non-singular $\Theta_{\rm Pl}$-bounce connecting a stage of ordinary contraction described by Einstein gravity with an anamorphic smoothing stage that involves modifications to Einstein gravity. 

Our $\Theta_{\rm Pl}$-bounce will involve a combination of modifications to the non-singular bounce described in Sec.~\ref{sec:recap}:
\begin{enumerate}
\item[--]a modified Hubble-like parameter $\Theta_m$ to track the evolution of the physical background;
\item[--] no effective Planck mass and, hence, no unique $\Theta_{\rm Pl}$ during the bounce stage but a new parameter $\gamma$ that smoothly approaches $\Theta_{\rm Pl}$ before and after the bounce;
\item[--] a new parameter (different than $-\dot{\gamma}/{\gamma^2}$) to track the evolution of stable curvature and tensor modes during the bounce stage;
\item[--] no relaxing of all non-trivial couplings after the bounce in order to smoothly join onto an anamorphic smoothing phase.
\end{enumerate}

During the brief $\Theta_{\rm Pl}$-bounce stage, the simple anamorphic action in Eq.~\eqref{JF-action1} is completed to form an ${\cal L}_4$-Horndeski,
\begin{align}
\label{JF-action-bounce}
S &= \int d^4x \sqrt{-g} \left( \frac{1}{2}\left(M_{\rm Pl}^0)^2(1+f(\phi) \right)R -  \frac{1}{2} K(\phi) (\partial\phi)^2 - \left(\frac{m_0}{m(\phi)}\right)^4V(\phi) \right) 
 \nonumber
\\
&+ \int d^4x \sqrt{-g} \left(q(\phi)X^2- b(\phi)X\Box\phi  \right) + S_{\rm matter}
\,.
\end{align}
Here, $f, q$ and $b$ are the specific Horndeski couplings as defined in the previous section and $K,V, m$ and $m_0$ are inherited from the anamorphic action in Eq.~\eqref{JF-action1}.
As we emphasized above, an important difference from simple generalized Brans-Dicke theories is the non-trivial `braiding' between gravity and the imperfect Horndeski scalar, which has the consequence that it is not possible to define a metric transformation and re-express the action such that the scalar is minimally-coupled to gravity, {\it i.e.,} the field theory is not conformally (or disformally) equivalent to Einstein gravity.
Consequently, the notion of `frame-invariants' and `effective Planck mass' has no true counterpart in the ${\cal L}_4$-Horndeski theory ({\it i.e.,} during the $\Theta_{\rm Pl}$-bounce stage) and is only well-defined whenever the ${\cal L}_3$-Horndeski coupling is turned off.  
Hence, instead of following $\Theta_m$ and $\Theta_{\rm Pl}$ through the phase governed by the ${\cal L}_4$-Horndeski, we note the conditions that have to be satisfied after the bounce phase is over and  the couplings $f, b, q$ are turned off:
\begin{itemize}
\item[{\it i.}] $\Theta_{\rm Pl}$ has to flip sign from negative to positive (`$\Theta_{\rm Pl}$-bounce');
\item[{\it ii.}] $\Theta_m$ has to keep its sign and remain negative (no $\Theta_m$-bounce). 
\end{itemize}

({\it i.}) {\it $\Theta_{\rm Pl}$-bounce:} Even though $M_{\rm Pl}$ and thus $\Theta_{\rm Pl}$ are not well-defined in the ${\cal L}_4$-Horndeski theory, the condition for $\Theta_{\rm Pl}$ to switch sign is unambiguous: the quantity $\gamma$ defined in Eq.~\eqref{gamma} must switch sign from negative to positive. This is due to the fact that before the Horndeski couplings $f, b,$ and $q$ are turned on, $\Theta_{\rm Pl}=H=\gamma<0$, and, identifying $1+f =M_{\rm Pl}^2$ after the cubic Horndeski couplings are turned off, $\Theta_{\rm Pl}=M_{\rm Pl}^{-1}(H+\dot{M}_{\rm Pl}/M_{\rm Pl})=M_{\rm Pl}^{-3}\gamma>0$. 

Note that, even though the sign flip of $\gamma$ is a sufficient condition for a successful $\Theta_{\rm Pl}$-bounce,  $\gamma$ does not determine  the spectra of linear cosmological perturbations (which are fixed by the more complex quantities $A, B$ given in Eqs.~(\ref{Aeq}-\ref{Beq})). Hence, $\gamma$ does not play the same physical role in the ${\cal L}_4$-Horndeski theory as $\Theta_{\rm Pl}$ does in generalized Brans-Dicke theories. 
Nevertheless, the dynamics of  co-moving curvature and tensor perturbations also gets modified as a result of the $\Theta_{\rm Pl}$-bounce:  before the $\Theta_{\rm Pl}$-bounce, both the scalar and tensor spectra are determined by the quantity $\alpha_m^3\epsilon_m \sim a^3$. Since $a$ is shrinking during contraction, the modes blue shift and decay. After the $\Theta_{\rm Pl}$-bounce, on the other hand, the modes' dynamics is determined by $\alpha_{\rm Pl}^3\epsilon_{\rm Pl} \sim \alpha_{\rm Pl}^3$, and, since $\alpha_{\rm Pl}$ is growing, the modes red shift and grow. 
%%%%%FIGURE~2
\begin{figure}[tb]
\begin{center}
\includegraphics[width=9cm]{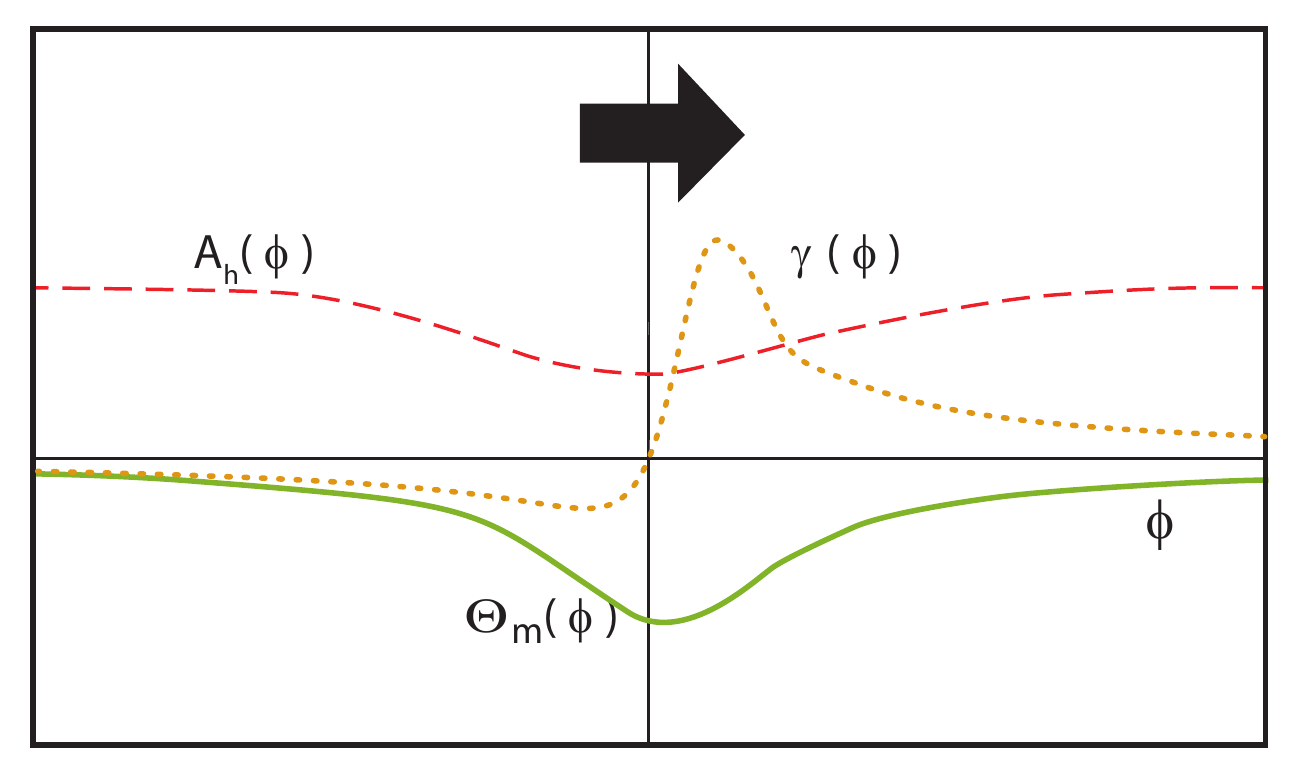}
\caption{
A schematic illustration of the Hubble parameter $\Theta_m$ (continuous green line), and the parameters $\gamma$ (dotted orange line) and $A_h$ (dashed red line)  as a function of the anamorphic field $\phi$ showing the evolution from the current dark-energy domination through the $\Theta_{\rm Pl}$-bounce stage (middle of the figure) up until the $\Theta_m$-bounce in the cyclic anamorphic scenario. Outside the $\Theta_{\rm Pl}$-bounce stage, $A_h\sim$ constant and $\dot{\gamma}<0$. $\Theta_{\rm Pl}$ is not plotted because, during the $\Theta_{\rm Pl}$-bounce stage, the effective Planck mass is not well-defined and so there is no way to define $\Theta_{\rm Pl}$. $\Theta_m$ is defined up to an overall normalization factor during the $\Theta_{\rm Pl}$-bounce. Here, we defined $\Theta_m$  such that $\Theta_m^2=H^2/(1+f)$. This definition ensures that $\Theta_m$ smoothly connects to its canonical definition in Eq.~\eqref{thetam} before and after the $\Theta_{\rm Pl}$-bounce. 
}
\label{FigIn}
\end{center}
\end{figure}
%%%%%%% 

({\it ii.}) {\it no $\Theta_m$-bounce:} Similar to finding a matching condition for $\Theta_{\rm Pl}$ switching sign, it is possible to define a sufficient condition for no $\Theta_m$-bounce occurring during the $\Theta_{\rm Pl}$-bounce. Although the effective Planck mass is not well-defined during this stage, this only means that the normalization of $\Theta_m\propto H+\dot{m}/m$ is ambiguous. The sign of $\Theta_m$ is well-defined  and determined solely by the sign of $H$. 
That is, $\Theta_m$ remains negative if $H$ is negative before, during, and after the $\Theta_{\rm Pl}$-bounce. A corollary is that no cosmological bounce occurs during the $\Theta_{\rm Pl}$-bounce.

A successful $\Theta_{\rm Pl}$-bounce is illustrated in Fig.~\ref{FigIn}.
We note that the only difference between the examples of Refs.~\cite{Ijjas:2016vtq,Ijjas:2017pei} reviewed in the previous section and the current one is that after the $\Theta_{\rm Pl}$-bounce the universe approaches an anamorphic stage rather than an Einstein gravity stage, {\it i.e.}, the frame-independent ratio of Planck mass and test-particle mass maintains time dependence with everything else remaining the same. This makes the construction even simpler. 
At the time of the $\Theta_{\rm Pl}$-bounce, the gravitational coupling $M_{\rm Pl}^2=1+f$  is beginning to increase and the kinetic coupling $k$ is beginning to decrease, as depicted in Fig.~\ref{FigCompleteScenario}.
It is at this point in the evolution that $m/M_{\rm Pl}$ first gain time dependence such that $\Theta_m$ and $\Theta_{\rm Pl}$ become distinguishable. 
In particular, $H$ and thus $\Theta_m$ remain finite and negative during and after the $\Theta_{\rm Pl}$-bounce, while $\Theta_{\rm Pl}$ becomes positive as a result of the sign flip of $\gamma$.
With increasing $f$, $\gamma$ switches sign at $t=t_{\gamma}$ while the quantity $A$ behaves as $ \sim \big(1+  (t-t_{\gamma})^2\big)$. Eventually,  the quantity $2k(\phi)+3(f,_{\phi})^2$  
becomes positive with the kinetic coupling $k$ remaining negative throughout.
Around this point, the couplings to higher order kinetic terms $q$ and $b$ (all functions of $\phi$) can be turned off such that the terms are negligible as $\phi$ increases further and the Lagrangian reduces to the anamorphic form in Eq.~\eqref{JF-action1}.

\section{Smoothing, $\Theta_m$-bounce, and transition to hot big-bang evolution} 
\label{sec:secExit}

The conditions are now precisely what is required for the anamorphic smoothing phase
(where $\phi_{\mathrm{ANA}} < \phi$):
A period of ordinary contraction drove the field uphill in $V$ and, through the brief period of the $\Theta_{\rm Pl}$-bounce stage, the sign of $\Theta_{\rm Pl}$ is flipped to be positive and opposite that of $\Theta_m$ so that the $\Theta_{\rm Pl}$-bounce stage ends at  approximately the value of $V(\phi)$ needed to generate energy-density perturbations of the right amplitude given by 
 \begin{align}
\label{amplitude}
\frac{\delta \rho}{\rho}\bigg|_{N=60} &= \sqrt{{\cal K}(\phi)}\sqrt{V(\phi)} \left(m_0^{-1}\frac{m}{M_{\rm Pl}}\right)^2
\left( \frac{d}{d\phi} \ln\left(\frac{m^4}{M_{\rm Pl}^4}V(\phi)\right) \right)^{-1} \bigg|_{N=60}\sim 10^{-5}
\,,
\end{align}
where the pre-factor ${\cal K}$ is defined as
 \begin{align}
\label{def-scriptK}
{\cal K}(\phi) &= \left(m_0^{-1}\frac{m}{M_{\rm Pl}}\right)^2\left( k(\phi)+6\left(m_0\frac{d}{d\phi}\left(\frac{m}{M_{\rm Pl}}\right)^{-1}\right)^2 \right)
\,.
\end{align}
In addition, as described above in Sec.~\ref{sec:secBasics}, for a phase of successful anamorphic smoothing, an array of conditions have to be satisfied. In the scalar-tensor realization of the anamorphic scenario defined through the Lagrangian in Eq.~\eqref{JF-action1}, these conditions correspond to the simple relation
\begin{equation}
\label{allconstraints}
0 < 3 + \frac{1}{2}k(\phi)\left(\frac{d}{d\phi} \left(\frac{ M_{\rm Pl}}{m}\right)\right)^2  < \epsilon  < 1\,.
\end{equation}
Here the left-hand inequality is the no-ghost condition, {\it i.e.}, the non-negativity condition for the pre-factor of the kinetic term $\sim\epsilon_{\rm Pl}\dot{\zeta}^2$ in the second-order action for co-moving curvature modes given in Eq.~\eqref{S2-0}, $\epsilon_{\rm Pl}= -\dot{\Theta}_{\rm Pl}/\Theta_{\rm Pl}^2= (1/2) {\cal K}(\phi)(\dot{\phi}/M_{\rm Pl})^2>0$.
The middle inequality is the condition for contraction ($\Theta_m<0$); and the third inequality is the smoothing and squeezing constraint ($\epsilon <1$). 

Due to the wrong-sign of the kinetic term $k$ the anamorphic field $\phi$ continues to move uphill along the potential energy curve $V$.  The gravitational coupling, $M_{\rm Pl}^2$ grows monotonically and so  the effective gravitational constant $\sim 1/M_{\rm Pl}^2$ becomes weaker, contributing to the anamorphic smoothing process.   
The Hubble-like parameters $\Theta_m<0$ and $\Theta_{\rm Pl}>0$ have opposite sign, though both $\Theta_m^2$ and $\Theta_{\rm Pl}^2$ are decreasing and very slowly approaching zero during the anamorphic phase while the mass-variation index $m/M_{\rm Pl}$ remains nearly constant. Notably, $\dot{\Theta}_m /M_{\rm Pl} > 0$ during the entire smoothing phase.  The very slow decrease of $\Theta_m^2$ and $\Theta_{\rm Pl}^2$  ensures that the curvature perturbations are (nearly) scale-invariant with a small red tilt. The scalar and tensor spectra can be computed as described in \cite{Ijjas:2015zma}, where the result depends on the choice of parameters for $V(\phi)$, $M_{\rm Pl}^2(\phi)$, and $k(\phi)$.

%%%%%FIGURE~3
\begin{figure}[tb]
\begin{center}
\includegraphics[width=8.5cm]{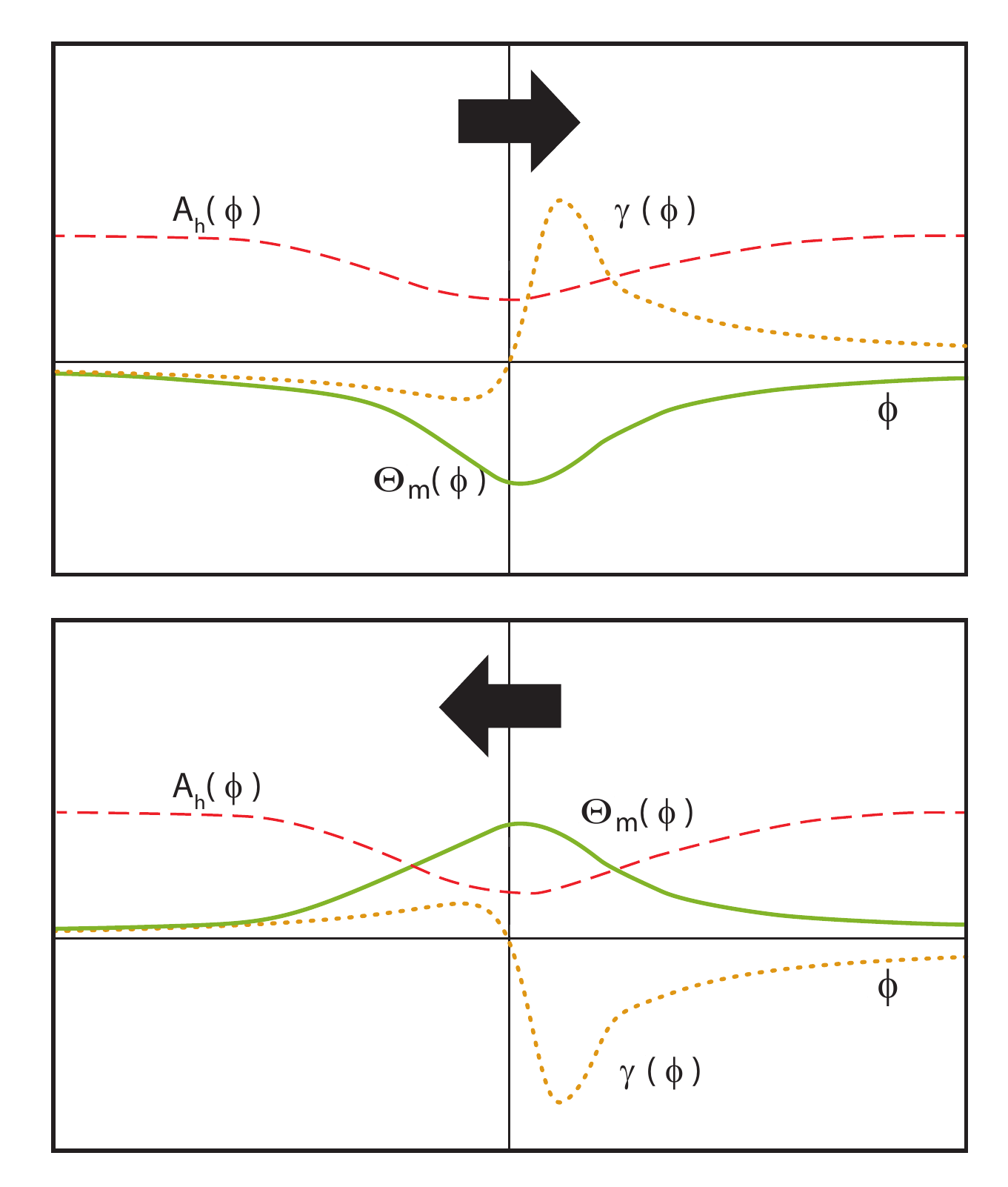}
\caption{
A schematic illustration of the Hubble-like parameter $\Theta_m$ (continuous green line), and the parameters $\gamma$ (dotted orange line) and $A_h$ (dashed red line)  as a function of the anamorphic field $\phi$ showing the transition to standard hot big bang evolution after the $\Theta_m$ bounce in the cyclic anamorphic scenario (lower figure) compared to the evolution from dark-energy domination to the $\Theta_m$-bounce (upper figure). 
}
\label{FigOut}
\end{center}
\end{figure}
%%%%%%% 

Anamorphic contraction ends when $\phi$ reaches infinity after time $\Delta t$, 
\begin{equation}
\label{timeTOinfty}
\Delta t\sim \int^{\infty}_{\phi_f} \frac{d\phi}{\dot{\phi}} \end{equation}
and $\Theta_m<0$ approaches zero.  It is straightforward to see that $\Delta t$ is finite for anamorphic smoothing: Since the background is contracting, the field's kinetic energy  $\sim\dot{\phi}^2$ is growing as the field rolls uphill; typically, $\dot{\phi} \sim \exp(p\phi)$, where $p$ is a constant, so that $\Delta t\sim (1/p) \exp(-p\phi_f)$. However, for $\Delta t$ to be finite, it is sufficient that the growth rate of $\dot{\phi}$ goes as $\phi^p$ with $p\geq2$. While $\dot{\phi}^2$ is growing, $\Theta_m$ continues to increase, which means it eventually switches from negative to positive, resulting in  a  $\Theta_m$-bounce. Note that the $\Theta_m$-bounce is the one and only cosmological bounce in the sense that this is the single event when the physical background transits from contraction to expansion.
The $\Theta_m$-bounce occurs at $\phi=\infty$, analogous to that in ekpyrotic models, in which $\dot{\phi}$ reverses sign and gets a small kick \cite{Bars:2013vba}.  
The only difference is that, when the field value reaches infinity, all physical quantities (such as $\Theta_m, \Theta_{\rm Pl}$, and the field?s kinetic and potential energies) remain finite. In this sense, the situation is simpler than the adjoint extension invoked in ekpyrotic models. 
Consequently, the universe starts expanding. No $\Theta_{\rm Pl}$-bounce occurs, though.  Rather $\dot{\Theta}_{\rm Pl}$ remains negative and  $\Theta_{\rm Pl}$  undergoes a simple reversal of sign: it is positive before the $\Theta_m$-bounce, reaches zero at the $\Theta_m$-bounce and continues to decrease and becomes negative after the $\Theta_m$-bounce.   
.

On the way downhill, the gravitational coupling $M_{\rm Pl}^2$ is shrinking and the kinetic coupling $k$ is growing, {\it i.e.,} becoming less negative. The field eventually re-enters the range  where the ${\cal L}_4$-Horndeski couplings turned on during the trajectory uphill. This occurs, when $6(m_0(M_{\rm Pl}/m),_{\phi})^2$ has become small again relative to $|k(\phi)|$. As before, $\dot{\gamma}>0$ leads to a $\Theta_{\rm Pl}$-bounce and $\Theta_{\rm Pl}$ becomes positive and no $\Theta_m$-bounce occurs.  As $\phi$ continues to decrease and  $\dot{M}_{\rm Pl} /M_{\rm Pl}$ vanishes, the invariants $\Theta_m, \Theta_{\rm Pl}$ (now both positive) as well as the frames become indistinguishable.  The field continues to roll down the potential energy curve, reaches the negative minimum and continues uphill. The small extra kick generated at the  $\Theta_m$-bounce enables the field to cross the tiny barrier and settle in the metastable vacuum state.  The anamorphic field oscillates around its minimum converting its remaining kinetic energy to radiation, thereby reheating the universe.  After 14 billion years of expansion and cooling, the universe reaches a condition like the present, dominated by the small, positive false vacuum energy density, and the cycle returns to stage 1. 
Note that, as in the ekpyrotic cyclic model \cite{Steinhardt:2002ih}, there is a long period of dark-energy domination such that fluctuations produced in previous cycles are redshifted to super-horizon wavelengths that are not visible in the next cycle.

\section{Summary and outlook}
\label{sec:secDiscussion}

Cyclic cosmologies involving epochs of expansion followed by  contraction have the advantage of resolving any sort of initial conditions problems related to assuming a beginning in time, such as the inflationary multiverse or self-reproduction problem. So far there has been only a single cyclic scenario, the classic cyclic ekpyrotic cosmology. 
In this paper, we have presented a second example by showing how the recently introduced anamorphic cosmology can be made cyclic.  

The key to the cyclic completion was the recent construction of a stable non-singular bounce. Remarkably, even though the bounce was originally developed to connect a period of ordinary contraction with  a period of expansion both described by Einstein gravity, here we showed how the construction can be adapted to connect a phase of ordinary contraction with an anamorphic smoothing phase that involves modifications to Einstein gravity.
Notably, the cyclic version of anamorphic cosmology does not violate the 2nd law of thermodynamics; as in the ekpyrotic case \cite{Steinhardt:2001st}, the average scale factor grows by an exponential factor cycle by cycle such that the entropy density associated with each cycle remains the same but the total volume and total entropy both grow proportionally.

The cyclic anamorphic scenario opens up many new avenues. In future work, we plan to explore how different versions  of cyclic anamorphosis based on different coupling functions and multiple fields affect the predictions and stability. We are also planning a fully non-perturbative treatment using methods of numerical relativity.

\section*{Acknowledgments}
We thank Paul J. Steinhardt for helpful discussions and comments on the manuscript. This research was supported by the Simons Foundation Program `Bouncing Cosmologies and Cosmological Bounces,'  as part of the `Origins of the Universe' Initiative.

\bibliographystyle{plain}
\bibliography{anamorphic2}

\end{document}